\documentclass[11pt,twoside]{article}

\usepackage{asp2006}
\usepackage{epsf}
\usepackage{psfig}
\usepackage{lscape}

\begin{document}
\title{The Automatic Galaxy Collision Software}
\author{Beverly J. Smith$^1$, Chris Carver$^1$, Phillip Pfeiffer$^2$,
Sam Perkins$^2$, Jason Barkanic$^2$, Steve Fritts$^2$, 
Derek Southerland$^2$, 
Dinikar Manchikalapudi$^2$, Matt Baker$^2$, John Luckey$^2$, 
Coral Franklin$^1$,
Amanda Moffett$^{1,3}$, and Curtis Struck$^4$}
\affil{$^1$Dept of Physics and Astronomy, East Tennessee State University.
$^2$Computer and Information Science Dept, East Tennessee State University.
$^3$Now at University of North Carolina, Chapel Hill.\\
$^4$Dept of Physics and Astronomy, Iowa State University.
}    

\begin{abstract} 
The key to understanding the physical processes that occur
during galaxy interactions
is dynamical modeling, and especially the detailed matching of numerical
models to specific systems.  To make modeling interacting galaxies
more efficient, we have constructed the `Automatic Galaxy Collision' (AGC)
code, which requires less human intervention in finding good matches
to data.  We present some preliminary results from this code for the
well-studied system Arp 284 (NGC 7714/5), and address questions of
uniqueness of solutions.
\end{abstract}


\section{Introduction}

Gravitational interactions and collisions between galaxies play an important
role in galaxy evolution.
Computer simulations 
of such encounters
can provide detailed matches to observations of
specific systems, providing information
about star formation triggering and mass transfer between galaxies
(\citealp{struck03}, hereafter SS03).
However, the process of finding the best
model for a particular system is time-consuming,
tedious, and subjective,
requiring a series
of runs with the parameters being varied by hand, until the
model matches the observed appearance of the galaxy, with a
good match being determined by eye.  
Thus, 
only a handful of galaxy pairs have been modeled in
detail.  
Furthermore, little information is provided
about the uncertainty in the parameters or the uniqueness of
the model.

To address these issues, we have created the `Automatic
Galaxy Collisions' (AGC) code.
This code combines a galaxy interaction code with a genetic algorithm
and a pattern-matching routine to automatically find models
that match a particular galaxy.
This modeling project is being done in parallel with a
Spitzer IR and GALEX UV imaging study
of three dozen nearby interacting galaxies 
\citep{smith07, smith10a, smith10b,
giroux10}.
Our goal is to use the AGC code to produce preliminary
models for our full set of galaxies.
These will then be used as
initial guesses for an N-body code, 
a method that was used in SS03. 

Our approach parallels that of
\citet{theis01}
and \citet{theis03},
who used a similar method to model the NGC 4449/DDO 125 and
M51 encounters.
Evolutionary codes have also been used to model 
the
M81/M82/NGC 3077 system
\citep{gomez04}
and the
Milky Way/LMC/SMC
interaction \citep{ruzicka06}.
An alternative approach is used by \citet{wallin10},
who are harnessing the efforts of Galaxy Zoo participants
to model galaxy interactions.
Another approach has been used by
\citet{barnes09},
whose
``Identikit'' 
interactive graphics program provides a way for 
users to find a model for a particular system.

\section{The Automatic Galaxy Collisions (AGC) Code }

Our AGC code combines a
standard restricted 3-body
galaxy interaction code \citep{wallin90} with a revised and updated
version of the PIKAIA
genetic algorithm 
\citep{charbonneau95}
and our own
pattern-matching routine.
The AGC code
runs a large number of galaxy interaction simulations
with randomly-selected interaction parameters.
It then uses the pattern-matching routine
to determine how well each model fits the observations
of the galaxy system.  Based on these fitnesses,
it then
runs a new generation of models, using parameters closer
to those of the real system.
Over many generations, one can converge upon
a model which matches the real system very well.
By using
multiple runs with different initial conditions,
the question of uniqueness can be investigated.

The original PIKAIA program was converted to a 
parallel code by \citet{met01}.
We have re-engineered this software, 
repackaging and streamlining the code,
improving the genetic algorithm,
making the code user-friendly, and creating a test suite
to validate the code.

The AGC pattern matching routine begins with a sky-subtracted 
smoothed
image of the galaxy 
system being modeled. 
Next, for each timestep of the simulation,
the model image is rotated 
to align
the galaxy-galaxy axis with that of the real 
system, scaled to match the radius of the 
primary galaxy, and regridded
to match the image.  Then, a
morphological chi-square is calculated for each timestep, 
summing over each pixel, weighting the tail and sky pixels differently
from the disk regions, and scaling down the model disk points 
to account
for dust extinction.
Finally,
a weighted chi-square is computed
to account for differences between the model and the real
kinematic line of nodes and
the separation between the two model galaxies compared
to that of the real galaxies.
The fitness is the 
reciprocal of 
this weighted chi-square.

\section{\bf Results on NGC 7714/5}

The AGC code was first tested using the interacting galaxy pair 
NGC 7714/5 (Figure 1a).
This system has been modeled 
`by hand'
using the \citet{wallin90} restricted 3-body code 
(\citealp{smith92}, hereafter SW92),
and an N-body code that 
includes gas hydrodynamics, star formation, and ISM heating 
(SS03). 

\begin{figure}
\centering
\plottwo{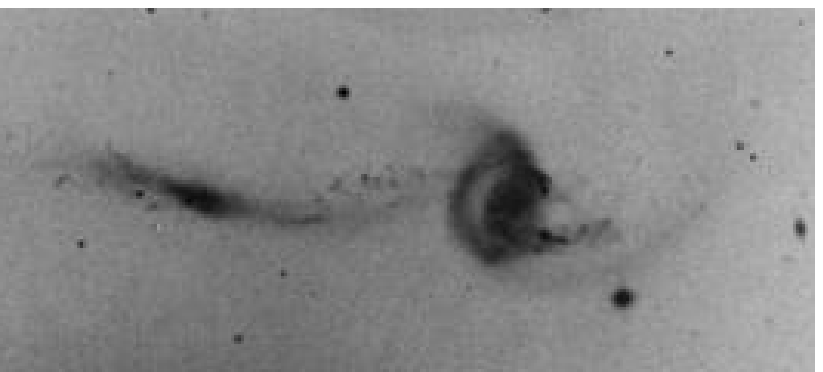}{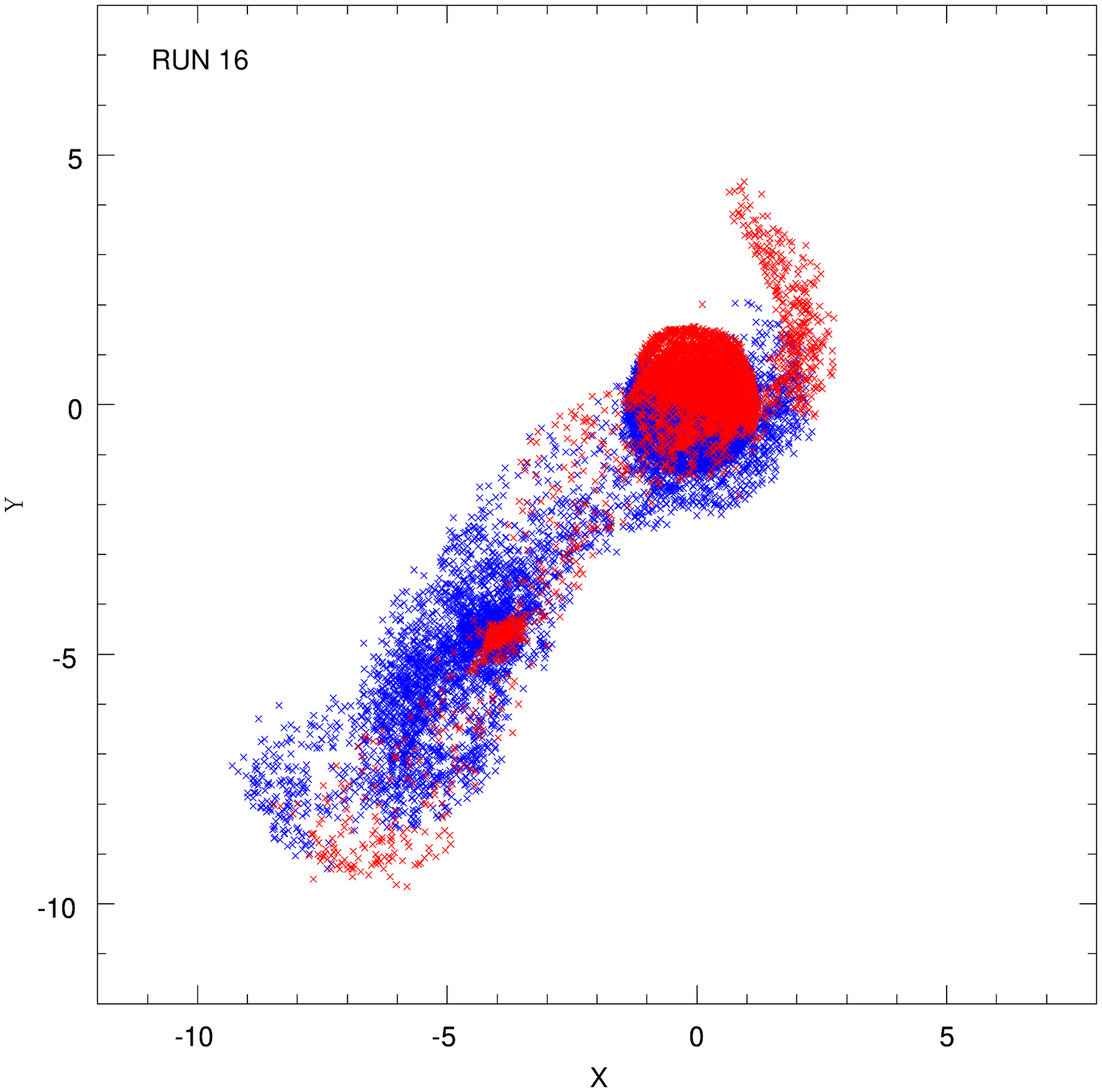}
\caption{Left:
Arp (1966) Atlas photograph of NGC 7714/7715 (Arp 284). North
is up while east is to the left. NGC 7714 is the larger galaxy to the west.
Right: The best-fit simulation for Run 16 - a Z model. 
}
\end{figure} 

Over 1.5 months, we did 
36 AGC runs for NGC 7714/5 with different initial random seed numbers.
These runs were made on the ETSU `blackpearl' computer cluster,
which consists of 60 Dell 1950 blade computers.
In these runs, only the 7 
parameters
that determine the positions, orientations, and orbits
of the two galaxies 
were allowed to vary.
The 36 runs converged to four different `best-fit' models.
Models W and X have retrograde orbits, while Y and Z are prograde.
Models W and Z are near-head-on collisions, while X and Y are impacts
near the edge of the disk.
All of these models approximately match the pair separation
and basic kinematics, and all do an adequate job of reproducing
the bridge.  To varying degrees, each model type also reproduces
the tidal tails, except for the X model, which does not match the
western NGC 7714 tails well.
The fact that different runs produced different `best' models 
shows
that multiple solutions are possible when
the input is limited to an optical image with line of node
information, and 
a restricted 3-body code is used.
However, the quality of the best-fit models and the limited
number of solutions 
shows that within these limitations the code works adequately,
and limits the set of candidate solutions to explore further.

Of the four model types, 
model Z has the highest fitness and the best match to the
system by eye.  The results of run 16, a Z model,
is shown in Figure 1b.
This does a good job of producing the bridge and the long western tail.
In 3-dimensional views, a long tail extends from the eastern
side of NGC 7715 
and
curves behind NGC 7714 away from the observer. This
is not visible `from Earth'
since it is in the same plane as the bridge and both galaxies.

\begin{figure}
\plottwo{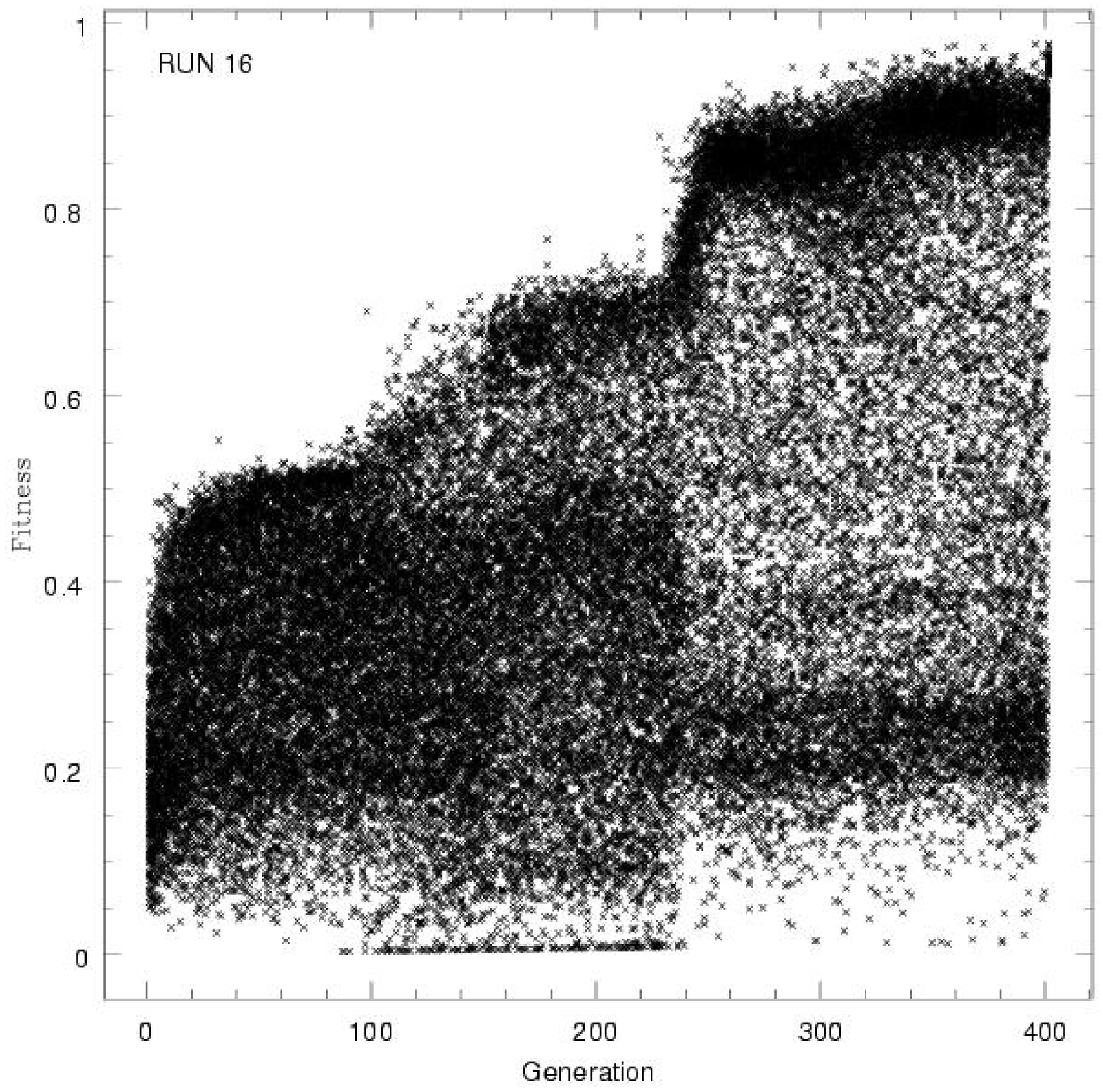}{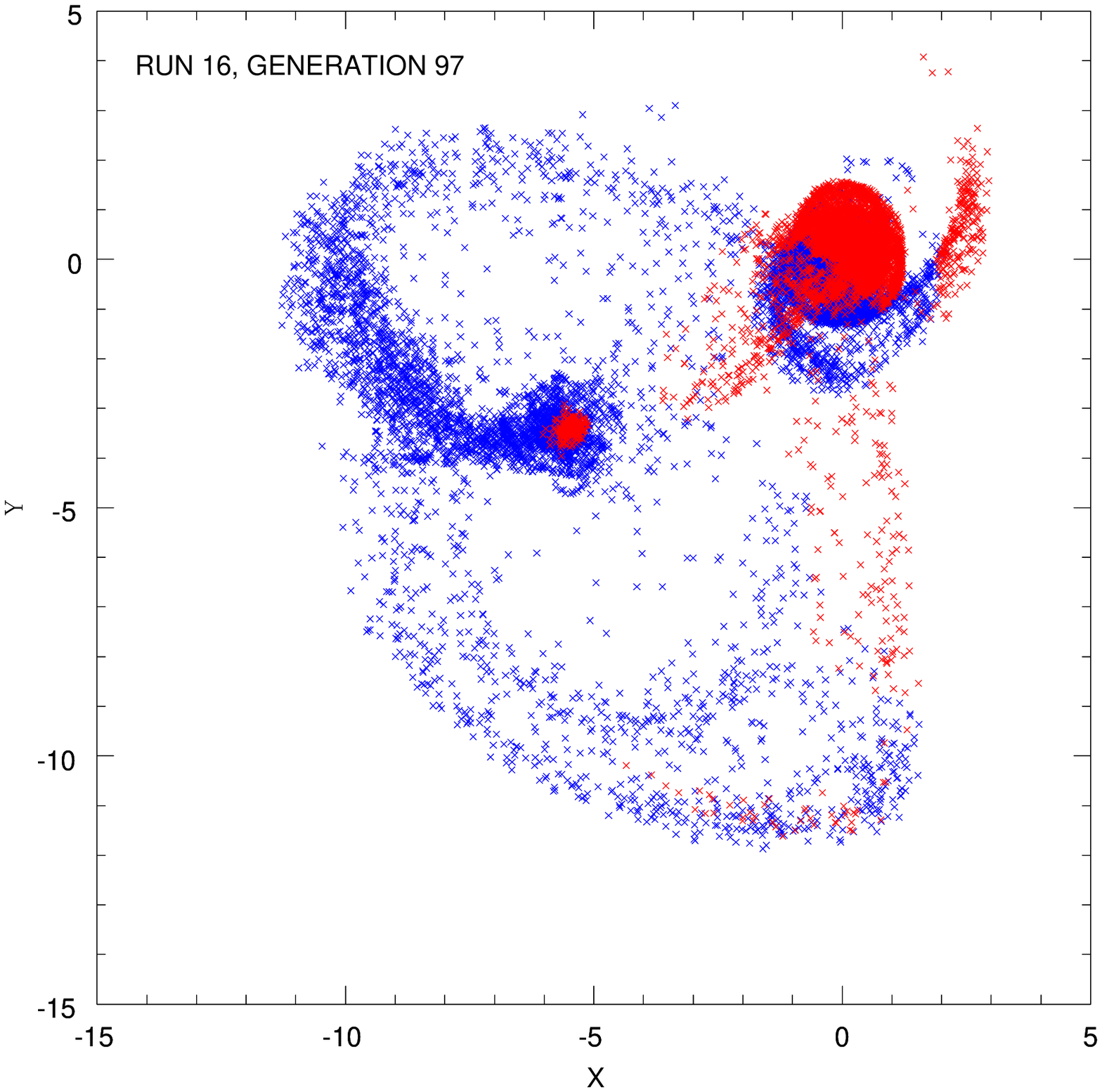}
\caption{
Left: Fitness plot for Run 16, showing the evolution of 
fitness with generation.
Right: The best-fit simulation in Generation 97 for Run 16.
}
\end{figure}

While the parameters of the earlier `by hand' SW92 model 
do not exactly match
any of our four model types, it exhibits 
similarities to our X model, being
a retrograde encounter with an off-center collision.
In contrast, the `by hand' SS03 hydrodynamical
model most closely resembles our Z model, being
a prograde near-head-on collision.
In particular, both the SS03 model and our Z model produce a long
eastern tail that curves back behind the bridge.

Figure 2a shows the corresponding fitness versus generation plot for run 16,
showing an improvement in the fitness with
increasing generation. 
After 61 generations, 
the best fit model has a large impact parameter and no tails. 
By generation 97, the best-fit model had
changed to an almost head-on collision. 
At this point, 
the western tail of NGC 7714 is reasonably well-matched,
but the NGC 7715 tails are incorrect (Figure 2b).
By the 259th generation, 
this discrepancy has been fixed,
and the tails are aligned with the bridge.

To further test our code, we took the best-fit model from Run 16
(Figure 1b) as the
`real' galaxy.
We then ran eight `rerun' runs, with this as our `input image'.
All 8 models converged to the same basic four model types.
This shows that, within the limitations of the code, we can 
reproduce the basic structure of the galaxy.   
Rerun 6 gave us the closest match, producing a Z model
that matches the Run 16 model 
in four of the seven parameters.
The main difference 
is the appearance of the NGC 7715 tails,
which are not lined up with the bridge.


\acknowledgements 
This research has been supported by NASA Long Term Space 
Astrophysics (LTSA) grant NAG5-13079 and East Tennessee 
State University Research Development grant RD0094.


\end{document}